\documentstyle[11pt]{article}

\def\fspace#1{\int_{\Gamma({#1},\theta)}\,dK}
\def\nocutint{\int\limits^y_0 dy'\,\gamma_0^2}

\def\ch{\mbox{ch}}
\def\sh{\mbox{sh}}
\begin{document}
\begin{flushright}
{FIAN/TD -21/95}
\end{flushright}
\begin{flushright}
{December 1995}
\end{flushright}
\bigskip
\begin{center}
{\bf{NONPERTURBATIVE DISSIPATION IN QCD JETS}}
\end{center}
\medskip
\begin{center}
{\bf{A.V.~Leonidov {\footnote{e-mail address leonidov@td.lpi.ac.ru}}
 and D.M.~ Ostrovsky {\footnote{e-mail address ostrov@td.lpi.ac.ru}}}}
\end{center}

\medskip
\begin{center}
{\it{Theoretical Physics Department, P.N.Lebedev Physics Institute
117924 Leninsky pr. 53, Moscow, Russia}}
\end{center}

\bigskip

\begin{center}
{\bf{Abstract}}
\end{center}

The phenomenological energy dissipation in QCD jet evolution taking
into account the colour coherence is considered and the corresponding
modified parton multiplicity and distribution function  are computed.

\newpage

\section{Introduction}

 Physics of QCD  jets is one of the most important testing grounds
for the theory of strong interactions. The perturbative evolution of  a
quark-gluon jet is at present well understood (see, e.g., \cite{BPQ}). The known
details include in particular the famous colour coherence phenomenon which
leads to a probabilistic picture of timelike jet evolution where the daughter
partons are emitted into a gradually shrinking cones.  The nonperturbative
aspects of a jet evolution are much less understood. From the experience
of a sum rule approach applied to the analysis of heavy resonance properties
 \cite{SVZ} it is clear, that when the virtuality of a daughter parton is
 of order of
1 {\mbox {GeV}} one has to take into account the nonperturbative corrections
due to the nonperturbative vacuum fields giving rise to the QCD vacuum
condensates.
Another situation in which one deals with dissipative effects is a jet
propagation in, e.g., nuclear medium. Although physically these situations
are very different, we expect the corresponding theoretical formalism
to be the same.

   In both cases the new interaction vertices lead to the appearance of new
dimensionful parameters. This means in turn that a scaling description
of jet evolution where the interaction vertices are determined only by
dimensionless quantities (energy or virtuality ratios) is no longer valid.
Physically what happens is a beginning of a string formation taking energy
from the perturbative component and converting it to nonperturbative  degrees
of freedom. The first model description of a jet evolution taking into account
the nonperturbative energy loss was proposed long ago by Dremin \cite{Dr}.
 It was
based on the analogy with the physics oh high energy electromagnetic showers
in the medium, where apart from a scale-invariant evolution due to a photon
bremmstrahlung and pair creation there appears a scale-noninvariant
energy dissipation due to the ionization of the atoms in the medium \cite{Be}.
Later the corresponding modified evolution equations were analytically solved
in \cite{DL1}, where the expressions for the parton multiplicities in quark and
 gluon
jets and the energy loss by the perturbative component were computed.
A discussion of this approach and also of the latest Monte-Carlo calculations,
\cite{GE} where the nonperturbative component was described by QCD effective
lagrangeans can be found in the recent review \cite{DL2}.  A serious drawback
 of Dremin equations \cite{Dr} is, however, a disregard of the kinematical
and color interference effects in the jet evolution which are known to be
of crucial importance for describing the jet characteristics. The main goal of
 the present paper is to introduce a formalism
providing a possibility of an analytical treatment of the jet evolution
taking into account both color interference and nonperturbative energy loss.

Before writing down the modified evolution equations let us recall, that color
coherence in QCD jets can be described using various approximations. The simplest
is a leading logarithmic one (DLA, see \cite{BPQ}). The calculations in this
approximation are relatively simple, but the energy of the jet is not conserved.
This is due to the eikonal description used in this approximation, when the
energy loss of the projectile is considered to be negligible. A more refined
analysis is made within a modified leading logarithmic approximation (MLLA),
where the total energy is conserved. The Dremin equations \cite{Dr}
correspond to a case, where the energy is perturbatively conserved. Thus to
calculate an absolute value of the dissipative energy loss one has to modify
an MLLA formalism. The corresponding equations can be written (see Appendix A),
but unfortunately we were not able to find their solutions. Thus in what
follows we shall concentrate on a simpler DLA case, where the energy
nonconservation is built in already at the level of perturbation theory.
In this case one can not calculate the absolute energy loss, but the relative
one is fairly well-defined.

\section{DLA Equation for the Generating Functional}

Let us turn to a description of a formalism which generalizes the known
effective methods of describing the QCD jet evolution including color
coherence effect \cite{BPQ} by providing a possibility of considering
the new possible nonperturbative sources of energy dissipation.

Technically it is most convenient to use a generating functional.
$G(\{u\})$  from which one can calculate both an exclusive cross section
\begin{equation}
d\sigma^{excl}_N=\left(\prod^N_{i=1}\,d^3k_i\,\frac{\delta}{\delta u(k_i)}
\right)\,G(\{u\})\Big|_{u=0}d\sigma_0.
\end{equation}
and an inclusive one
\begin{equation}
d\sigma^{incl}_N=\left(\prod^N_{i=1}\,d^3k_i\,\frac{\delta}{\delta u(k_i)}
\right)\,G(\{u\})\Big|_{u=1}d\sigma_0.
\end{equation}
In the following we shall use the notations
$dK=\gamma_0^2\frac{dk}{k}\,\frac{d^2k_\perp}{2\pi k^2_\perp}$ for the
usual DLA integration measure and  $\Gamma(p,\theta)$ for the possible gluon
emission phase space domain, where $p$ is an initial gluon energy and
$\theta$ is an opening angle of a jet.  Let us remind, that in DLA
approximation  the generating functional $G(p,\theta;\{u\})$ satisfies
the following master equation

\begin{flushleft}
$G(p,\theta;\{u\}) = u(p)e^{-w(p,\theta)}+$
\end{flushleft}
\begin{equation}\label{Master}
\int_{\Gamma(p,\theta)} {dk \over k}
{d^2 k_{\perp} \over {{k}_{\perp}}^2} {2 C_V \alpha_s \over \pi}
e^{-w(p,\theta)+w(p,\theta_k)} G(p,\theta_k;\{u\}) G(k,\theta_k;\{u\}) ,
\end{equation}
where
$$
 w(p,\theta)= \int_{\Gamma(p,\theta)} dK
$$
is a total probability of an  emission of a gluon having an energy $p$
into the cone having an opening angle $\theta$, and $\exp (-w(p,\theta))$
is a formfactor giving a probability of a non-emission of the same gluon.
Then a sandard procedure (\cite{BPQ}) leads to the following equation
on the generating functional:
\begin{equation}\label{Gold}
G(p,\theta;\{u\})=u(p)\,exp\left(\fspace{p}\,[G(k,\theta_k;\{u\})-1]\right).
\end{equation}

Let us now discuss the possible nonperturbative modifications of the equations
for the generating functional. The aim is to describe a partial convertion
of the perturbative degrees of freedom (gluons) and their energy
into the nonperturbative
ones by the process presumably analogous to the ionization in the case of
electromagnetic showers discussed in the Introduction. Physically the
picture for the jet is that the dissipation pumps some energy from the
cone corresponding to the perturbative jet evolution reducing the gluon
multiplicity and energy inside it. Thus we have some additional nonperturbative
vertex coupling the perturbative gluons with the nonperturbative vacuum or
nuclear medium.
The presence of such a coupling can be related to reducing a probability of an
emission of a perturbative gluon by taking away a certain portion of its energy
before the perturbative emission. Techically it is convenient to introduce a
following modification of the gluon emission probability:
\begin{equation}
 w(p,\theta;\beta)= \int_{\Gamma(p-\beta,\theta)} dK,
\end{equation}
where the constant $\beta$ describes the additional damping of a perturbative
 gluon emission due to a dissipative interaction with the nonperturbative
vacuum fluctuations or nuclear medium introduced by compressing a  possible
 phase space domain
for the gluon emission from $\Gamma(p,\theta)$ to $\Gamma(p-\beta,\theta)$.
 The corresponding modification of the master equation
(\ref{Master}) reads
\begin{flushleft}
$G(p,\theta;\{u\}) = u(p)e^{-w(p,\theta;\beta)}+$
\end{flushleft}
\begin{equation}
\int_{\Gamma(p-\beta,\theta)} {dk \over k}
{d^2 k_{\perp} \over {{k}_{\perp}}^2} {2 C_V \alpha_s \over \pi}
e^{-w(p,\theta;\beta)+w(p,\theta_k;\beta)}
G(p,\theta_k;\{u\}) G(k,\theta_k;\{u\}) ,
\end{equation}
Then instead of the Eq.~(\ref{Gold}) we get
\begin{equation}\label{Gnew}
G(p,\theta;\{u\})=u(p)\,exp\left(\fspace{p-\beta}\,[G(k,\theta_k;\{u\})-1]
\right),
\end{equation}
where the initial conditions for this equation read $G(p,\theta;\{u\})|_{u=1}=1$.

We suggest that $\beta \ll p$ for all physically meaningful values of energy
and in the following we shall  consider the terms of the first order in the
 small parameter $\varepsilon=\beta/p$. In Appendix B we illustrate
the thus arising perturbation theory by deriving the differential equation on
 the
leading correction to the generating functional.

\section{Mean multiplicity}

Let us begin this section by looking at the simplest jet characteristic, a
mean multiplicity of partons in a jet (in the following we are considering only
gluons). Following the standard steps we obtain a following integral equation
on the mean multiplicity ${\bar{n}}(p,\theta)$:
\begin{equation}
{\bar{n}}(p,\theta)=1+\fspace{p-\beta}\,{\bar{n}}(k,\theta_k).
\end{equation}\label{phas}
In the following it will be convenient to rewrite the integration over the phase
space as
\begin{equation}
\fspace{p-\beta}=\int_{\Gamma(p-\beta,\theta)}\frac{dk}{k}\,
\frac{d^2k_\perp}{2\pi k^2_\perp}\gamma_0^2=
\int\limits^{y-\varepsilon}_0 d\xi\,\int\limits_0^\xi dy'\gamma_0^2,
\end{equation}
where
$$
y=\ln\frac{p\theta}{Q_0},
y'=\ln\frac{k\theta_k}{Q_0},
\xi=\ln\frac{k\theta}{Q_0}.
$$
Working in the first order in $\varepsilon$ we can write
$$
{\bar{n}} (p,\theta)={\bar{n}}_0 (y)+\varepsilon  {\bar{n}}_1 (y).
$$
The equations for the functions ${\bar{n}}_0(y)$ and ${\bar{n}}_1(y)$
 take the form
\begin{equation}
{\bar{n}}_0 (y)=1+\nocutint\, {\bar{n}}_0 (y')
\end{equation}
and
\begin{equation}
 {\bar{n}}_1 (y)=
\nocutint\, {\bar{n}}_1 (y')\,(e^{y-y'}-1)\,-\,\nocutint\, {\bar{n}}_0 (y')
\end{equation}
The first one has a well known solution
$$
{\bar{n}}_0 (y)= \ch(\gamma_0 y),
$$
and the second one can be rewritten in the differential form
\begin{equation}\label{dif}
{\bar{n}}_1 ''(y)- {\bar{n}}_1'(y)
=\gamma_0^2 {\bar{n}}_1 (y)+\gamma_0^2(\ch(\gamma_0y)
-\gamma_0\sh(\gamma_0 y))
\end{equation}
with the initial conditions
$ {\bar{n}}(0)=0 ,  {\bar{n}}'(0)=-\gamma_0^2.$
The solution reads
\begin{equation}
\bar n_1=\gamma_0^2\ch(\gamma_0y)-\gamma_0\sh(\gamma_0y)+
+\gamma_0\frac{\lambda_2e^{\lambda_1y}-\lambda_1e^{\lambda_2y}}
{\sqrt{1+4\gamma_0^2}}.
\end{equation}
In the limit of $(y\gg1,\gamma_0\ll1)$ we have
\begin{equation}\label{n1}
{\bar{n}}\approx\frac{1}{2}e^{\gamma_0y}(1-\varepsilon\gamma_0-
2\varepsilon\gamma_0^3e^y)=
\frac{1}{2}\left({\frac{Q}{Q_0}}\right)^{\gamma_0}\left(1-\gamma_0
\frac{\beta}{p}-2\gamma_0^3\frac{\beta\theta}{Q_0}\right)
\end{equation}
The above result is in accord with the expectation based on physical reasoning.
Namely, the dissipation should lead to a decrease in perturbative multiplicity.
Let us note, that from Eq. (\ref{n1}) it is clear, that in order for
perturbation theory to be applicable one should have
$\beta \ll p/\gamma_0$ and $\beta \ll Q_0/\gamma_0^3$. This means that apart
from the expected expansion parameter $\varepsilon$  there appears a new one
$\varepsilon'= 2\gamma_0^2\frac{\beta\theta}{Q_0}$. An interesting feature
of the above answer is a multiplicative combination of the dissipation scale
$\beta$ and a perturbative coupling constant hidden in $\gamma_0$.

\section {Energy distrubution of particles in a jet}

Let us now turn to a computation of a one-particle energy distribution function
$\bar D(k,\theta)$. It can be obtained from the generating functional in the
following way:
\begin{equation}
\bar D(k,\theta)=k\frac{\delta}{\delta u(k)}G(p,\theta;\{u\})\Big|_{u=1},
\end{equation}
Making use of (\ref{Gnew}) and (\ref{phas}) we get
\begin{equation}\label{en}
\bar D(l,y)=\delta(l)+\int\limits^{l-\varepsilon}_0 dl' \int\limits^y_0 dy'
\gamma_0^2 \bar D(l',y'),
\end{equation}
where $l=\ln(p/k)=\ln(1/x)$ and $l'=\ln(k'/k)$. The integration is performed
over $k'$ and $\theta_{k'}$. Let us stress that in the above equation (\ref{en})
the distrubution function $\bar D$ is also $\beta$ - dependent. Expanding
it in the small parameter $\varepsilon$
$$
\bar D(l,y)=\bar D_0(l,y)+\varepsilon \bar D_1(l,y)
$$
we rewrite (\ref{en}) in the form
\begin{equation}
\bar D_0(l,y)+\varepsilon\bar D_1(l,y)=\delta(l)+
\int\limits^{l-\varepsilon}_0 dl' \int\limits^y_0 dy'\gamma_0^2
(\bar D_0(l,y)+\varepsilon e^{l-l'}\bar D_1(l,y)).
\end{equation}
Considering the zero and first order in $\varepsilon$ we have
\begin{equation}\label{D0eq}
\bar D_0(l,y)=\delta(l)+\int\limits^{l}_0 dl' \int\limits^y_0 dy'
\gamma_0^2 \bar D_0(l',y').
\end{equation}
\begin{equation}\label{D1eq}
\bar D_1(l,y)=\int\limits^{l}_0 dl' \int\limits^y_0 dy'\gamma_0^2 e^{l-l'}
\bar D_1(l',y')-\int^y_0 dy'\gamma_0^2 \bar D_0(l,y').
\end{equation}
The solution of equation (\ref{D0eq}) reads
\begin{equation}
\bar D_0(l,y)=\delta(l)+\gamma_0\sqrt{\frac{y}{l}}I_1(2\gamma_0\sqrt{yl})
\end{equation}
and that for $\bar D_1$ is (for details see Appendix C):
\begin{equation}\label{ensol}
\bar D_1(l,y)=-y\gamma_0^2\delta(l)-\gamma_0^2(e^l+1)\frac{y}{l}
I_2(2\gamma_0\sqrt{yl})+\gamma_0^3\left(\frac{y}{l}\right)^{3/2}(e^l-1)
I_3(2\gamma_0\sqrt{yl}).
\end{equation}
The plot of the resulting distribution for the jet energy $20 {\mbox { GeV}}$
is given at Fig.~1. We see that, as anticipated, the effect of dissipation
shows itself through the noticeable reduction of distribution function.
Let us also mention that here we also have a multiplicative dependence on both
perturbative and nonperturbative factors.

\section{Conclusions}

In this paper we have proposed a phenomenological nonperturbative
modification of the equations describing the evolution of QCD jets
and taking into account the crucial feature of the colour coherence
of the QCD cascades
by accounting for nonperturbative energy dissipation. The corresponding
modification of the MLLA and DLA formalism exploiting the generating
 functional was proposed. The calculation of simplest jet characteristics
such as mean multiplicity and energy spectrum of particles in a jet in a
DLA approximation has demonstrated an expected decrease in  multiplicity
 and corresponding changes in the energy distribution.

 An interesting feature of the result is an unusual
perturbative damping of the introduced nonperturbative energy dissipation
appearing in the multiplicative dependance on some power of the QCD coupling
constant times the dissipative scale.
It is tempting to relate this feature with the successes of the soft
blanshing hypothesis, where it is assumed that the nonperturbative
effects are not crucially essential for the jet characteristics. It is
interesting to see, whether such damping is present within a more realistic
MLLA description. Work in this direction is in progress.

\section{Acknowledgements}
We are grateful to I.M.~Dremin and I.V.~Andreev for the useful discussions.
A.L. is grateful to P.V.~Ruuskanen for kind hospitality in the University
of Helsinki, where part of this work was done. The research was supported by
Russian Fund for Fundamental Research, Grant 93-02-3815.

\newpage
\def\newappendix{
\section{}
\centerline{\Large\it Appendix \Alph{section} }
\bigskip}
\setcounter{section}{0}
\setcounter{equation}{0}
\def\thesection{  }
\def\theequation{\Alph{section}.\arabic{equation}}

\newappendix

Let us start with the proposed general modification of the equation for
the generating functional (we are closely following  the standard notations
 of \cite{BPQ}):
\begin{eqnarray}
G_A(p,\theta ;\{u\}) &=&u(p)\ e^{-w_A(p,\theta )}+\int_{\Gamma (p,\theta
)}dK\ \gamma _0^2(k_{\bot })\ e^{-w_A(p,\theta )+w_A(p,\theta _k)}*
\nonumber   \\
&&\ast \left[ \frac 12K_A^{BC}(\frac kp)\ G_B(k,\theta _k;\{u\})\
G_C(p-k,\theta _k;\{u\})-\right.  \label{start} \\
&&\ \ \quad \left. -L_A^B(k,p)\ \ G_B(k,\theta _k;\{u\})\right] .  \nonumber
\end{eqnarray}
 Here $L^B_A(k,p)$ is a phenomenologically introduced vertex responsible for
 the nonperturbative energy dissipation. It is convenient to rewrite the
above equation in the form:
\begin{eqnarray}
G_A(p,\theta ;\{u\})\ e^{w_A(p,\theta )} &=&u(p)+\int_{\Gamma (p,\theta
)}dK\ \gamma _0^2(k_{\bot })\ \ e^{w_A(p,\theta _k)}*
\nonumber   \\
&&\ast \left[ \frac 12K_A^{BC}(\frac kp)\ G_B(k,\theta _k;\{u\})\
G_C(p-k,\theta _k;\{u\})-\right.  \label{restart}\\
&&\ \ \quad \left. -L_A^B(k,p)\ \ G_B(k,\theta _k;\{u\})\right] .   \nonumber
\end{eqnarray}
This equation has to be supplied by the boundary condition $G_A=1$ at $u=1$.
This gives an equation of the formfactor $w_A(p,\theta)$:

\[
\ e^{w_A(p,\theta )}=1+\int_{\Gamma (p,\theta )}dK\ \gamma _0^2(k_{\bot })\
\ e^{w_A(p,\theta _k)}\ \left[ \frac 12\sum_{B,C}K_A^{BC}(\frac
kp)-\sum_BL_A^B(k,p)\right] ,
\]
where $z=\frac{k}{p}$. Standard procedure leads us to the following
 master equation:
\begin{eqnarray}
\frac{\partial G_A(p,\theta )}{\partial \ln \theta } &=&\int\limits_0^1dz\
\gamma _0^2(k_{\bot })\ \left[ \frac 12K_A^{BC}(z)\ \left\{ G_B(zp,\theta )\
G_C((1-z)p,\theta )-\sum_{B,C}G_A(p,\theta )\right\} -\right.
\nonumber \label{master}\\
&&\quad \qquad \qquad \quad \quad \left. -L_A^B(z\cdot p,p)\left\{
G_B(zp,\theta )\ -\sum_BG_A(p,\theta )\right\} \right]
\end{eqnarray}

Let us now consider a pure qluodynamics and a specific form of the
nonperturbative vertex
 $L_A^B(z\cdot p,p)=\ \nu \ \delta (1-\frac{\beta ^{\prime }}p-z)$. Then the
 master equation takes the form

\begin{eqnarray*}
\frac{\partial G}{\partial \ln \theta } &=&\int\limits_0^1dz\ \gamma
_0^2(k_{\bot })\ \left[ \frac 12K(z)\ \left\{ G(zp,\theta )\ G((1-z)p,\theta
)-G(p,\theta )\right\} \right] - \\
&&\qquad \qquad \quad -\nu \ \gamma _0^2(k_{\bot })\left\{ G(p-\beta
^{\prime },\theta )\ -G(p,\theta )\ \right\}
\end{eqnarray*}

In the interesting case of $\beta\ll p$  we get
\[
G(p-\beta ^{\prime },\theta )\ -G(p,\theta )\ \approx -\beta ^{\prime }\frac{%
\partial G}{\partial p}
\]
and

\begin{equation}\label{d/dE}
\frac{\partial G}{\partial \ln \theta }=\int\limits_0^1dz\ \gamma
_0^2(k_{\bot })\ \left[ \frac 12K(z)\ \left\{ G(zp,\theta )\ G((1-z)p,\theta
)-G(p,\theta )\right\} \right] +\beta ^{\prime }\nu \gamma _0^2\frac{%
\partial G}{\partial p}.
\end{equation}

The above equation contains a characteristic derivative over energy (the last
term). This term is responsible for the energy dissipation and has previously
been used for describing the dissipative energy losses in the
electromagnetic showers \cite{Be} and QCD cascades \cite{Dr},
\cite{DL1}.

\section{}
\centerline{\Large\it Appendix \Alph{section} }
\setcounter{equation}{0}
\bigskip

In this Appendix we derive an equation for the first correction to the
generating functional. Let us start with the basic equation
\begin{equation}\label{init}
G(p,\theta;\{u\})=u(p)\,exp\left(\fspace{p-\beta}\,[G(k,\theta_k;\{u\})-1]
\right)
\end{equation}
with the initial condition $G(p,\theta;\{u\})|_{u=1}=1$. Expanding the
generating functional we get in the zeorth and first orders in  $\varepsilon$:
\begin{equation}
G(p,\theta;\{u\})=G_0(y;u)+\varepsilon G_1(y;u),
\end{equation}
where $y=ln(p\theta/Q_0)$, $G_0(y;1)=1,G_1(y;1)=0$ and $\varepsilon=\beta/p.$

An argument of the exponent in (\ref{init}) is
$$
\nocutint(y-y')(G_0(y';u)-1)-\varepsilon\nocutint(G_0(y';u-1))-
$$
\begin{equation}
-\varepsilon\nocutint(e^{y-y'})G_1(y';u).
\end{equation}

After expanding this exponent to the first order in $\varepsilon$ we arrive at
\begin{equation}\label{inter1}
G_0(y;u)=u\,exp\left[\nocutint(y-y')(G_0(y';u)-1)\right]
\end{equation}
and
\begin{equation}\label{inter2}
G_1(y;u)=G_0(y;u)\left[\nocutint(e^{y-y'}-1)G_1(y';u)-
\nocutint(G_0(y';u)-1)\right],
\end{equation}
or in a simpler form
\begin{equation}
{G_0}'(y)=G_0(y)\nocutint(G_0(y')-1)
\end{equation}
and
\begin{equation}
G_1(y)=G_0(y)\nocutint(e^{y-y'}-1)G_1(y')\,-{G_0}'(y)
\end{equation}

The further simplification can be achieved by introducing a new function $H(y)$
by
\begin{equation}
H(y)=\int\limits^y_0 dy'(e^{y-y'}-1)G_1(y')\quad {G_1}'(y)=H'{'}(y)-H'(y)
\end{equation}
After the simple calculation we obtain the following differential equation
on the leading correction to the generating functional
\begin{equation}
H'{'}(y)-H'(y)=\gamma_0^2 G_0(y)H(y)-{G_0}'(y)
\end{equation}
The solutions of the above equation could be used as a starting point
for the analysis of the
dissipative corrections to the scaling behaviour of the moments and, in general,
to the KNO property of the leading approximation.

\section{}
\centerline{\Large\it Appendix \Alph{section} }
\setcounter{equation}{0}
\bigskip

In this Appendix we shall give some details on the computation of the
correction to the distribution function $\bar D(l,y)$. We start with the
equaitons  Eqs. (18), (19):
$$
\bar D(l,y)=\bar D_0(l,y)+\varepsilon \bar D_1(l,y)
$$
\begin{equation}
\bar D_0(l,y)=\delta(l)+\int\limits^{l}_0 dl' \int\limits^y_0 dy'
\gamma_0^2 \bar D_0(l',y').
\end{equation}
\begin{equation}
\bar D_1(l,y)=\int\limits^{l}_0 dl' \int\limits^y_0 dy'\gamma_0^2 e^{l-l'}
\bar D_1(l',y')-\int^y_0 dy'\gamma_0^2 \bar D_0(l,y').
\end{equation}

The solution of equation (\ref{D0eq}) is performed by  substituting
$$
\bar D_0(l,y)=\delta(l)+\sum C^{(0)}_{m,n} l^m y^n,
$$
giving
$$
C^{(0)}_{m,n}=\delta_{m,m+1}\frac{\gamma_0^{2(m+1)}}{m!(m+1)!}
$$
and
\begin{equation}
\bar D_0(l,y)=\delta(l)+\gamma_0\sqrt{\frac{y}{l}}I_1(2\gamma_0\sqrt{yl})
\end{equation}
Let us look for the solution for $\bar D_1$ in the form
$$
\bar D_1=-y\gamma_0^2\delta(l)+e^{l}\sum C^{(1)}_{m,n} l^m y^n.
$$
The solution for $C^{(1)}$ reads
\begin{equation}
C^{(1)}_{m,n}=
\left\{
\begin{array}{ccl}
-(-1)^{m+n}\frac{\gamma_0^{2n}}{n!m!}\sum\limits_{p=0}^{n-2}
C_{m-p}^{m-n+2}\,+\delta_{m+2,n}\left(-\frac{\gamma_0^{2(m+1)}}{m!n!}\right)&
\,&n\leq m+2\\
0&\,&{\rm otherwise}
\end{array}
\right.
\end{equation}
Thus we obtain the correction for the energy distribution
\begin{equation}
\bar D_1(l,y)=-y\gamma_0^2\delta(l)-e^{l}\left[\frac{y}{l}
I_2(2\gamma_0\sqrt{yl})+\sum\limits_{m,n}(-)^{m+n}\frac{\gamma_0^{2n}}{m!n!}
\sum\limits^{n-2}_{p=0}C_{m-p}^{m-n+2} l^m y^n \right],
\end{equation}
which can finally be be written in a compact analytical form
Eq.~(\ref{ensol}):
\begin{equation}
\bar D_1(l,y)=-y\gamma_0^2\delta(l)-\gamma_0^2(e^l+1)\frac{y}{l}
I_2(2\gamma_0\sqrt{yl})+\gamma_0^3\left(\frac{y}{l}\right)^{3/2}(e^l-1)
I_3(2\gamma_0\sqrt{yl}).
\end{equation}

\end{document}